\documentclass{jpp}
\usepackage{graphicx}

\usepackage[utf8]{inputenc}
\usepackage[T1]{fontenc}
\usepackage{amsmath}
\usepackage{units}

\shorttitle{Upgrade of the positron system of the ASACUSA-Cusp experiment}
\shortauthor{The ASACUSA-Cusp Collaboration}

\title{Upgrade of the positron system of the ASACUSA-Cusp experiment}

\author{
A.~Lanz\aff{1,2}
\corresp{\email{andreas.lanz@oeaw.ac.at}},
C.~Amsler\aff{1},
H.~Breuker\aff{3},  
M.~Bumbar\aff{1},
S. Chesnevskaya\aff{1},
G. Costantini\aff{4},
R. Ferragut\aff{5}, 
M. Giammarchi\aff{6},
A. Gligorova\aff{1}, 
G. Gosta\aff{4}, 
H.~Higaki\aff{7}, 
E.~D.~Hunter\aff{1},
C.~Killian\aff{1}, 
V.~Kraxberger\aff{1,2},
N.~Kuroda\aff{9}, 
M.~Leali\aff{4},
G.~Maero\aff{10},
C.~Mal\-bru\-not\aff{11}\footnote{present address: TRIUMF, Vancouver, Canada}, 
V.~Mascagna\aff{4},
Y.~Matsuda\aff{9},  
V.~M{\"a}ckel$^{a}$\footnote{present address: INFICON GmbH, K\"oln},
S.~Migliorati\aff{4}, 
D.~J.~Murtagh\aff{1} ,
A.~Nanda\aff{1,2},  
L.~Nowak\aff{1,2,11},
F.~Parnefjord~Gustafsson\aff{1,11},
S.~Rheinfrank\aff{1},
M.~Romé\aff{10},
M.~C.~Simon\aff{1},
M.~Tajima\aff{8}\footnote{present address: Japan Synchrotron Radiation Research Institute, Hyogo, Japan},  
V. Toso\aff{5}, 
U.~Uggerh{\o}j\aff{12},
S.~Ulmer\aff{3},  
L.~Venturelli\aff{4},
A.~Weiser\aff{1,2},
E.~Widmann\aff{1},
Y.~Yamazaki\aff{3}, 
\and 
J.~Zmeskal\aff{1}.
}

\affiliation{
\aff{1}Stefan Meyer Institute, Vienna, Austria
\aff{2}University of Vienna, Vienna Doctoral School in Physics, Vienna, Austria
\aff{3}Ulmer Fundamental Symmetries Laboratory, RIKEN, Saitama, Japan
\aff{4}Dipartimento di Ingegneria dell'In\-formazione, Universit\`a degli Studi di Brescia
and INFN Pavia, Italy 
\aff{5}Politecnico di Milano, Milano, Italy
\aff{6}INFN Milano, Milano, Italy
\aff{7}Graduate School of Advanced Science and Engineering, Hiroshima University, Hiroshima, Japan
\aff{8}Nishina Center for Accelerator-Based Science, RIKEN, Saitama, Japan
\aff{9}Institute of Physics, the University of Tokyo, Tokyo, Japan
\aff{10}Dipartimento di Fisica, Università degli Studi di Milano and INFN Milano, Italy
\aff{11}Experimental Physics Department, CERN, Geneva, Switzerland
\aff{12}Department of Physics and Astronomy, Aarhus University, Aarhus, Denmark
}   
\begin{document}

\maketitle

\begin{abstract}
The ASACUSA-Cusp collaboration has recently upgraded the positron system to improve the production of antihydrogen. Previously, the experiment suffered from contamination of the vacuum in the antihydrogen production trap due to the transfer of positrons from the high pressure region of a buffer gas trap. This contamination reduced the lifetime of antiprotons. By adding a new positron accumulator and therefore decreasing the number of transfer cycles, the contamination of the vacuum has been reduced. Further to this, a new rare gas moderator and buffer gas trap, previously used at Aarhus University, were installed. Measurements from Aarhus suggested that the number of positrons could be increased by a factor of four in comparison to the old system used at CERN. This would mean a reduction of the time needed for accumulating a sufficient number of positrons (of the order of a few million) for an antihydrogen production cycle. Initial tests have shown that the new system yields a comparable number of positrons to the old system.
\end{abstract}

\section{Introduction}
\label{sec:Introduction}
The ASACUSA-Cusp collaboration aims to measure the hyperfine splitting of a spin-polarised, ground-state antihydrogen beam in a magnetic field free region with a relative precision of parts per million \citep{WIDMANN200431}. Antihydrogen atoms are produced by mixing positrons and antiprotons, primarily via three-body recombination, in the so-called Cusp trap (due to its cusped magnetic field) \citep{Mohri_2003}. The highly inhomogeneous magnetic field focuses the two low-field seeking states on-axis while defocusing the high-field seeking states, yielding a spin-polarised beam of antihydrogen at the spectroscopy line. The first antihydrogen was successfully produced by \cite{Enomoto2010}. Antihydrogen was then observed \unit[2.7]{m} from the production region by \cite{Kuroda2014}. A subsequent measurement of the quantum state distribution at the position of the microwave cavity showed that the majority of antihydrogen atoms are in Rydberg states and the observed rate is too low for performing a spectroscopy measurement \citep{Kolbinger2021}. Since then, the focus of experiments is to increase the production of ground-state antihydrogen, which depends strongly on the density and temperature of the positron plasma \citep{Murtagh2014}. The first step was the upgrade of the Cusp trap in 2021, leading to a significant decrease of the temperature of an electron plasma \citep{Hunter2022}. The second step was the upgrade of the positron system to increase the number available for antihydrogen production and to reduce contamination of the ultra-high vacuum (UHV) of the Cusp trap during transfer.

In this work, the apparatus of the new system is described as well as the initial tests after the installation into the ASACUSA experimental area. In Section~\ref{sec:Experimental setup}, technical details of the moderator system, the buffer gas trap and the accumulator are given, followed by the results of the individual systems in Section~\ref{sec:Results} and a comparison with the previous system in Section~\ref{sec:discussion}.

\section{Positron System}
\label{sec:Experimental setup}
\begin{figure}
  \centering
  \includegraphics[width=\linewidth]{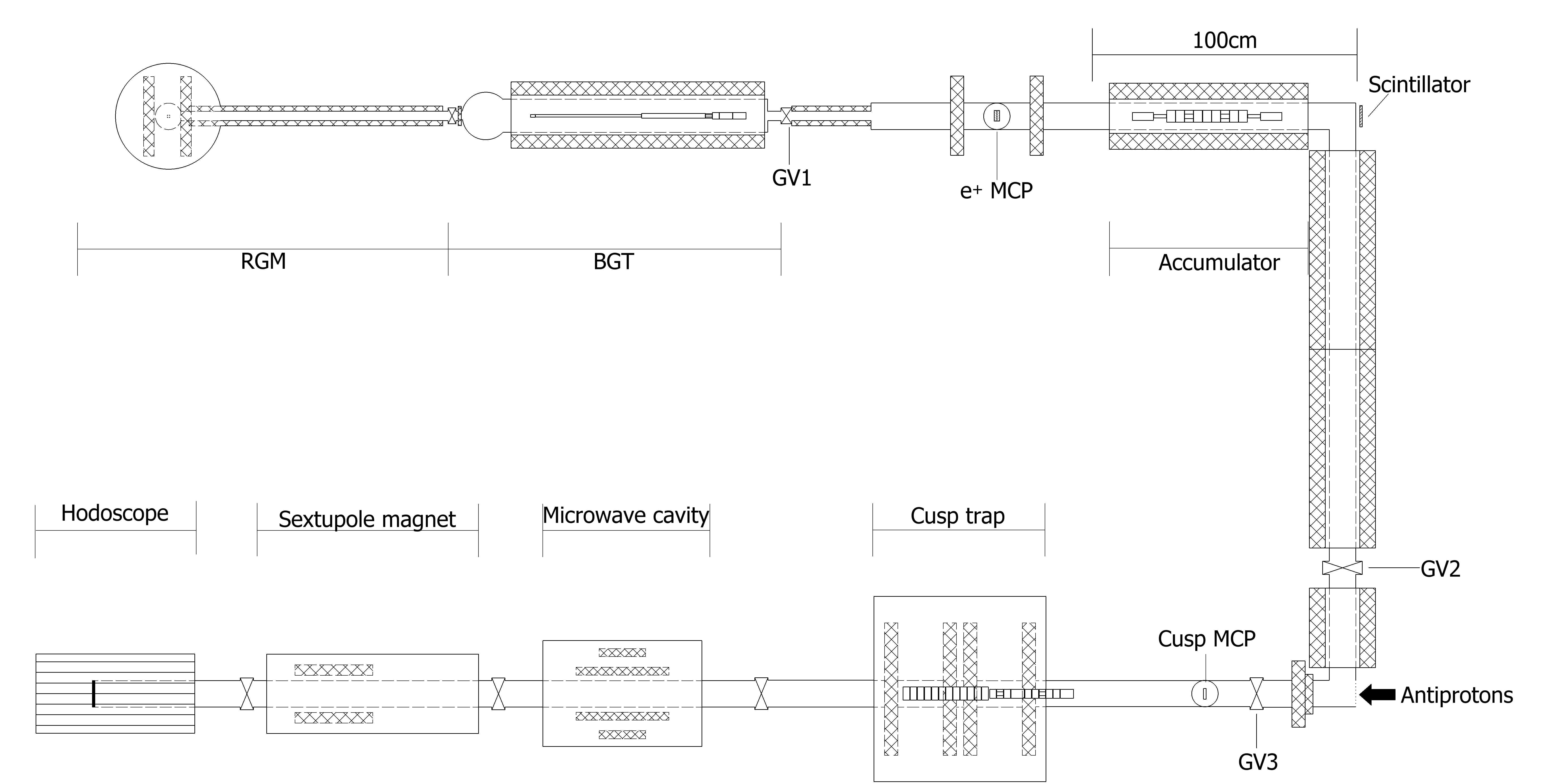}
  \caption{Scale drawing of the experimental setup excluding the antiproton apparatus. Positrons from the RGM and BGT are accumulated in the accumulator and then transferred to the Cusp trap. The individual apparatuses are labeled as well as the MCP detectors and the plastic scintillator. The magnets are indicated with cross hatching (the sizes of the solenoid magnets along the beamline are magnified for visibility), the gate valves are indicated and the three most important for positron accumulation and transfer are labelled with GV1, GV2 and GV3. The trap electrodes are shown in their position in the magnets.}
\label{fig:experiment}
\end{figure}

The new positron system consists of three parts: a commercial rare-gas moderator (RGM) and buffer gas trap (BGT) from First Point Scientific Inc. (FPS), which was previously used at Aarhus University, and an accumulator. The position of the traps, magnets, gate valves, and detectors (microchannel plate (MCP) and plastic scintillator) are indicated in the scale drawing of the experimental setup shown in Fig.~\ref{fig:experiment}.

\subsection{Rare Gas Moderator \& Buffer Gas Trap}
\begin{figure}
  \centering
  \includegraphics[width=\linewidth]{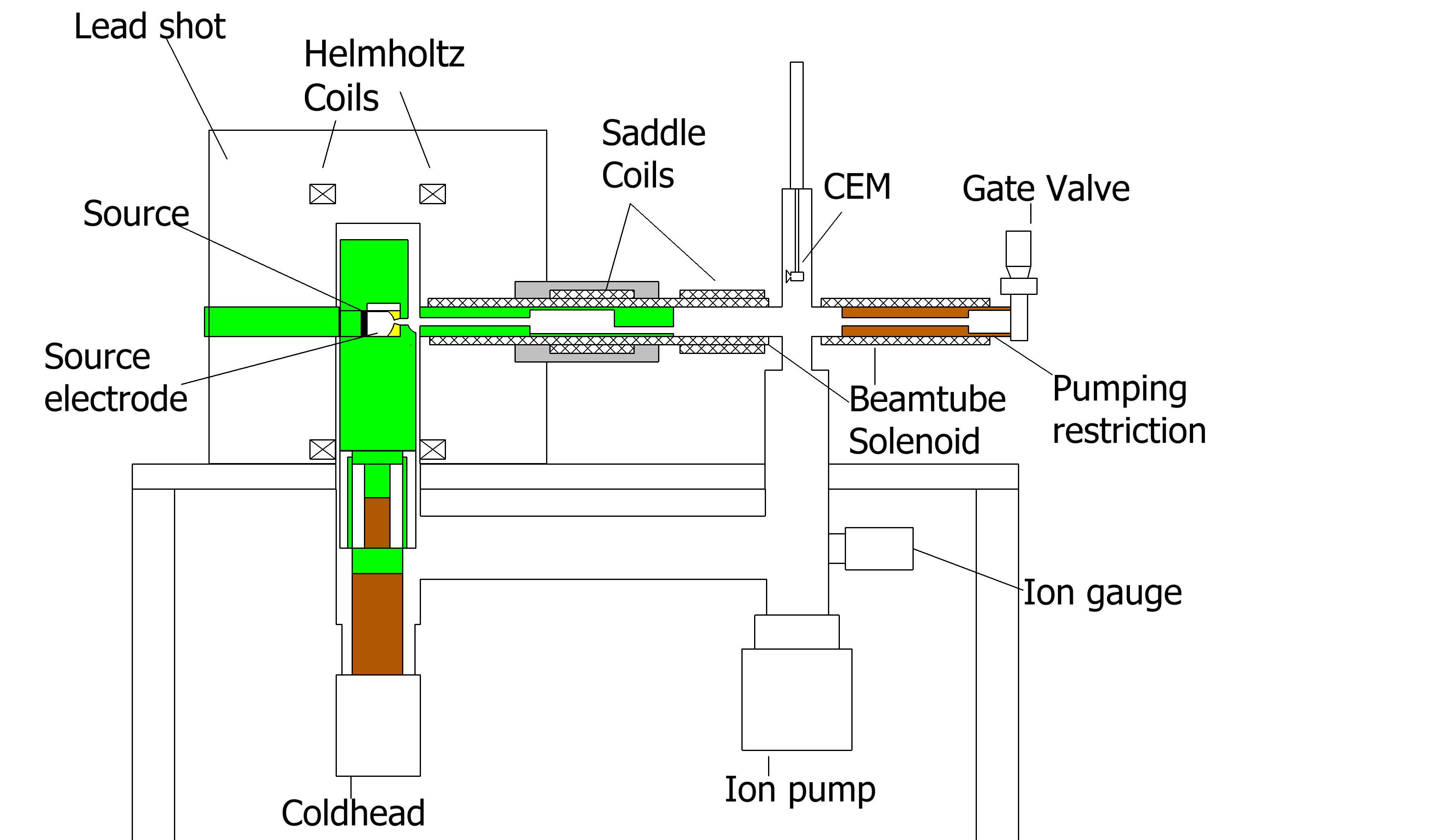}
  \caption{Schematic drawing of a cross-section of the RGM. The elkonite shielding is indicated in green, copper in brown, lead in grey and the magnets with cross hatching.}
\label{fig:rgm}
\end{figure}

The RGM and the BGT comprise a standard system for producing bunches of positrons. In this section only a short overview of the system is given. More details on the operation of the RGM and the positron traps can be found elsewhere \citep{Surko1989, Murphy1992, Mills1986}.

A schematic drawing of the RGM system is shown in Fig.~\ref{fig:rgm}. Positrons are produced by a commercial $^{22}$Na source from iThemba labs, purchased in 2011 with an activity of \unit[1.89]{GBq}. The source is housed in an elkonite shielded, cone-shaped electrode mounted on a cryocooler. The high-energy positrons from the source are moderated by a Ne-ice moderator \citep{Mills1986} which produces a slow beam with an energy of tens of eV, depending on the bias of the electrode. The positrons are magnetically guided from the source (\unit[120]{G}) through the beamtube solenoid (\unit[250]{G}) and focused with the matching coil (\unit [175]{G}) into the BGT (\unit[750]{G}). To prevent unmoderated positrons from reaching the trap, a \unit[30]{cm} long elkonite rod is inserted in the beamline which has an inner diameter of \unit[0.8]{cm} and is offset by \unit[1]{cm}. Two saddle coils producing a perpendicular magnetic field (\unit[23]{G}) give the moderated positrons a vertical offset and then realign them on axis, while unmoderated positrons annihilate on the rod. Additionally, this setup serves as a biological shield. When no positrons are required for accumulation, no current is applied to the saddle coils in which case positrons annihilate on the rod.

\begin{figure}
  \centering
  \includegraphics[width=\linewidth]{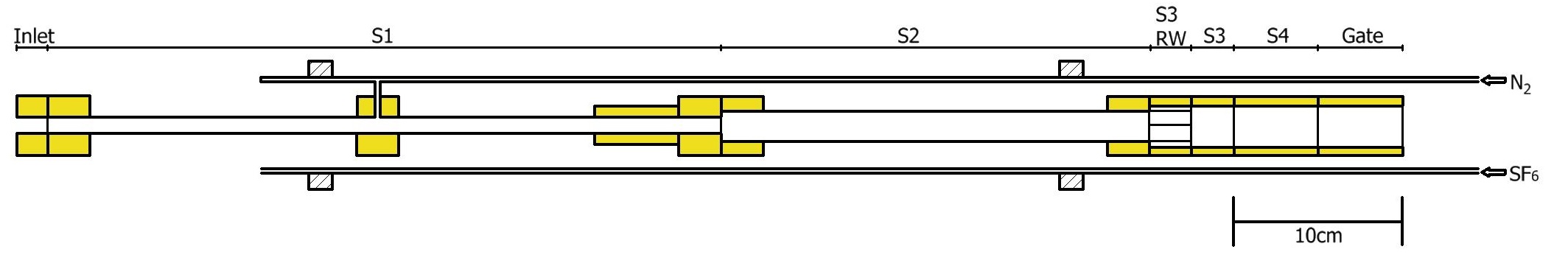}
  \caption{Scale drawing of the BGT electrodes. The slow beam enters from the left. N$_2$ is introduced in S1, while SF$_6$ is introduced into the chamber.}
\label{fig:BGT}
\end{figure}

After moderation, the positrons are trapped in the BGT. A scale drawing of the electrode structure is shown in Fig.~\ref{fig:BGT}. It consists of seven electrodes: the Inlet (length (l): \unit[18.5]{mm}, inner diameter (ID): \unit[10]{mm}), S1 (l: \unit[400]{mm}, ID: \unit[10]{mm}), S2 (l: \unit[255]{mm}, ID: \unit[18]{mm}), S3-RW (l: \unit[25]{mm}, ID: \unit[18]{mm}), S3 (l: \unit[25]{mm}, ID: \unit[25]{mm}), S4 (l: \unit[50]{mm}, ID: \unit[25]{mm}) and the Gate electrode (l: \unit[50]{mm}, ID: \unit[25]{mm}). N$_2$ gas is introduced into S1 producing a pressure gradient from S1 (order of \unit[$10^{-3}$]{mbar}) to the trapping region (order of \unit[$10^{-6}$]{mbar}). The given pressures in the trap were simulated using Molflow+ \citep{kersevan_introduction_2009} and agree with the calculation of the pressure in the trapping region using the measured lifetime. After multiple collisions with the N$_2$ molecules, those positrons which have not formed positronium have lost enough energy that they can no longer escape from the potential well produced in the lower pressure region by S3-RW, S3, S4 and Gate electrodes. For cooling the positrons SF$_6$ is introduced into the vacuum chamber \citep{Greaves2000} and a rotating wall (RW) electric field is applied to the eight-fold split electrode S3-RW (opposing electrodes are connected) to counteract radial expansion \citep{isaa:11}. After the fill time and a short cooling phase\textemdash typically after about one second\textemdash the positrons are pulsed out of the trap using a fast high voltage pulser connected to the gate electrode. 

The setup used at CERN is a modified version of the system as used at Aarhus University \citep{Andersen2015}: The fast high voltage amplifiers from the original system have been changed to slow, low noise amplifiers, and RC filters have been added to reduce the noise reaching the electrodes. The waveform generator for the RW has been replaced by BK Precision 4054b arbitrary waveform generators. To have better control over the trap potentials, the S3 and the RW electrodes, which were initially connected, have been separated to apply individual voltages on those two electrodes. Additionally, the control of the trap voltages has been upgraded to allow the possibility of introducing additional potential manipulations for moving the positrons.

\subsection{Accumulator}
\begin{figure}
  \centering
  \includegraphics[width=\linewidth]{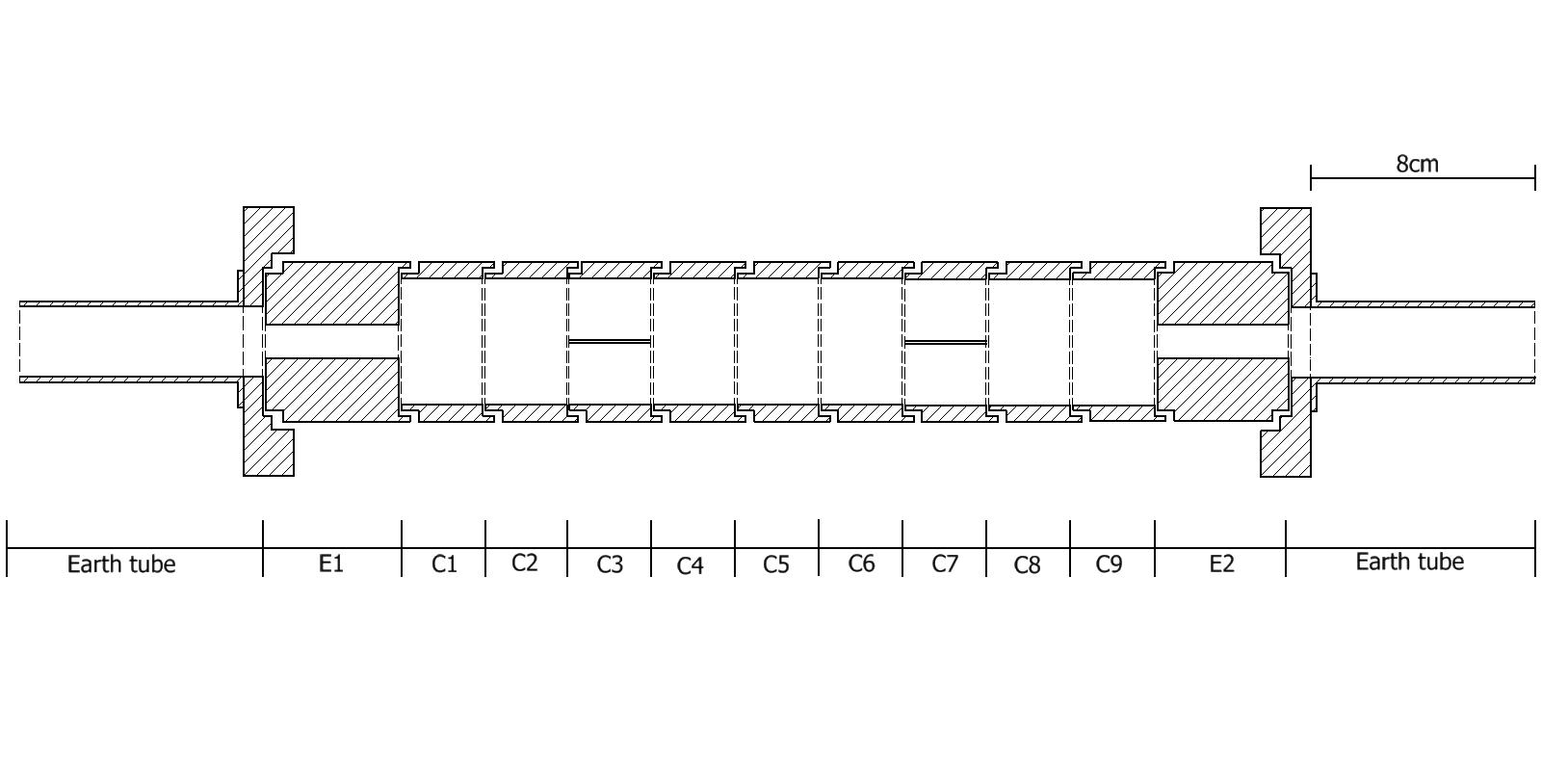}
  \caption{Scale drawing of the accumulator electrodes. Positrons from the BGT come from the left. C3 and C7 are split electrodes for the application of a RW. The electrodes are spray-painted with colloidal graphite and are separated by \unit[3]{mm} thick ceramic spacers, which are not shown in the drawing.}
\label{fig:stacker}
\end{figure}

The accumulator is a Penning-Malmberg trap housed in a \unit[1100]{G} field produced by a solenoid magnet, designed to accumulate several bunches (or "stacks") of positrons coming from the BGT. A scale drawing of the trap structure is shown in Fig.~\ref{fig:stacker}, which consists of eleven aluminium electrodes: E1 (l: \unit[47.5]{mm}, ID: \unit[12]{mm}), C1-C9 (l: \unit[29]{mm}, ID: \unit[45]{mm}) and E2 (l: \unit[47]{mm}, ID: \unit[12]{mm}) which are spray-painted with colloidal graphite and separated by \unit[3]{mm} thick ceramic spacer rings. C3 and C7 are four-fold split electrodes to apply a RW, E1 and E2 have smaller diameters to act as pumping restrictions and are designed to be used as pulsed electrodes to catch and extract positrons. The positron bunches from the BGT are magnetically guided into the accumulator by a \unit[30]{cm} long solenoid magnet (\unit[470]{G}) and two coils mounted in a Helmholtz configuration around the MCP (central field \unit[210]{G}). The pressure downstream of the accumulator is \unit[$2.0\cdot10^{-7}$]{mbar} during accumulation, due to the high pressure of the trapping and cooling gas from the FPS. This setup provides enough gas to cool the positrons in the accumulator, such that no additional cooling gas supply is needed. After accumulation is finished, the gate valve between the BGT and the accumulator, labelled as GV1 in Fig.~\ref{fig:experiment}, is closed, the gas is pumped out and the transfer of positrons into the Cusp trap is performed. 

\section{Results}
\label{sec:Results}
The results from commissioning of the RGM, the BGT and the accumulator are shown in Table~\ref{tab:results}. The measurements of the individual properties are described in more detail in the following subsections.

\begin{table}
  \begin{center}
\def~{\hphantom{0}}
  \begin{tabular}{lcc}
      Property  & RGM \& BGT & Accumulator \\ 
      Moderator efficiency (\%) & $0.2\pm0.1$ & -- \\ 
      Trapping efficiency (\%) & $15\pm4$ & -- \\
      Transfer efficiency (\%) & -- & $92^{+8}_{-11}$ \\
      Lifetime (s) & $3.9\pm0.8$ & $104 \pm 22$ / > 600 \\ 
      Energy spread (eV) & $1.49\pm0.01$ & -- \\ 
      Radius (mm) & $2.12\pm0.03$  & $1.44\pm0.05$-$1.85 \pm0.06$ \\ 
      Expansion rate (mm/s) & $2.24\pm0.08$ & $0.045\pm0.003$ / $0.013\pm0.004$ \\ 

  \end{tabular}
  \caption{Results of the operation of the positron system at CERN. For the accumulator, lifetime and expansion rate are reported for GV1 open/closed. The energy spread of the accumulator was not measured. The range of the radius in the accumulator corresponds to a range of \unit[5-160]{stacks}. The expansion rate was measured using \unit[30]{ stacks} and an initial positron cloud radius of \unit[$(1.63\pm0.04$)]{mm}.}
  \label{tab:results}
  \end{center}
\end{table}

\subsection{Rare Gas Moderator \& Buffer Gas Trap}
The activity of the source was \unit[89]{MBq} at the time the data were taken at CERN. 

The moderator efficiency is measured by counting the moderated positrons with a channel electron multiplier (CEM, model: KBL15RS/90-EDR from Dr. Sjuts Optotechnik GmbH) which can be inserted into the beamtube (see Fig.~\ref{fig:rgm}). The detection efficiency at a bias voltage of \unit[$-2.1$]{kV} is estimated, using the log-normal fit function with the parameters given in \cite{Newson2022}, to be \unit[$ (40 \pm 7)$]{\%}. Thus, the measured count rate, $55000\,\text{events/s}$, equates to $(1.4 \pm 0.3)\cdot 10^5$ slow positrons per second, or a moderator efficiency of \unit[$(0.2 \pm 0.1)$]{\%}, which is a factor of two to three smaller than previously reported for this system \citep{Andersen2015}. The efficiency may be slightly higher because we have not accounted for possible annihilations on the $79\%$ transparency, grounded mesh attached to the front of the CEM holder. The electric field produced by the \unit[$-2.1$]{kV} bias would, however, tend to focus the positrons in between the mesh-tines, increasing the effective transparency.  

\begin{figure}
  \centering
  \includegraphics[width=0.47\linewidth]{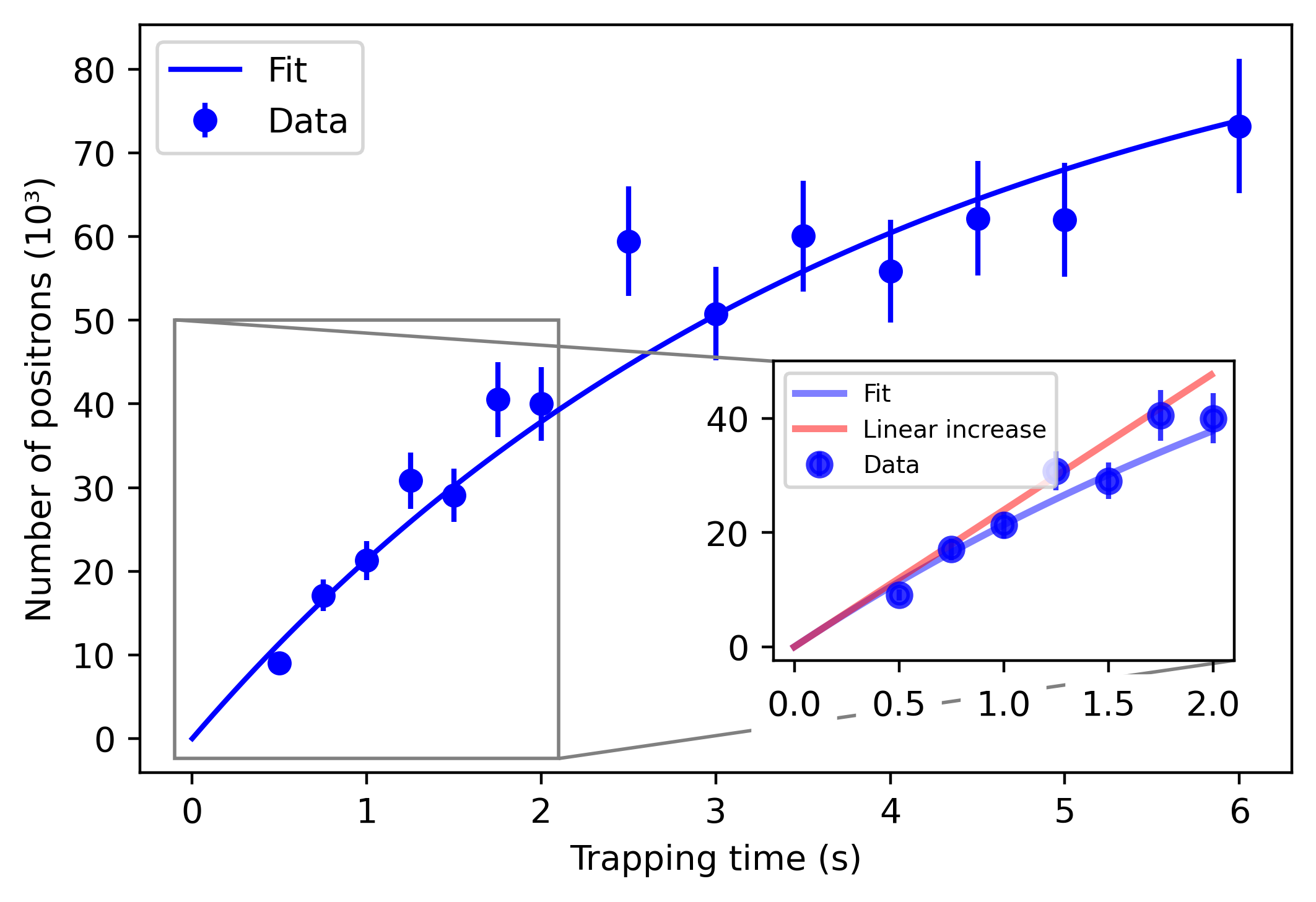}\includegraphics[width=0.53\linewidth]{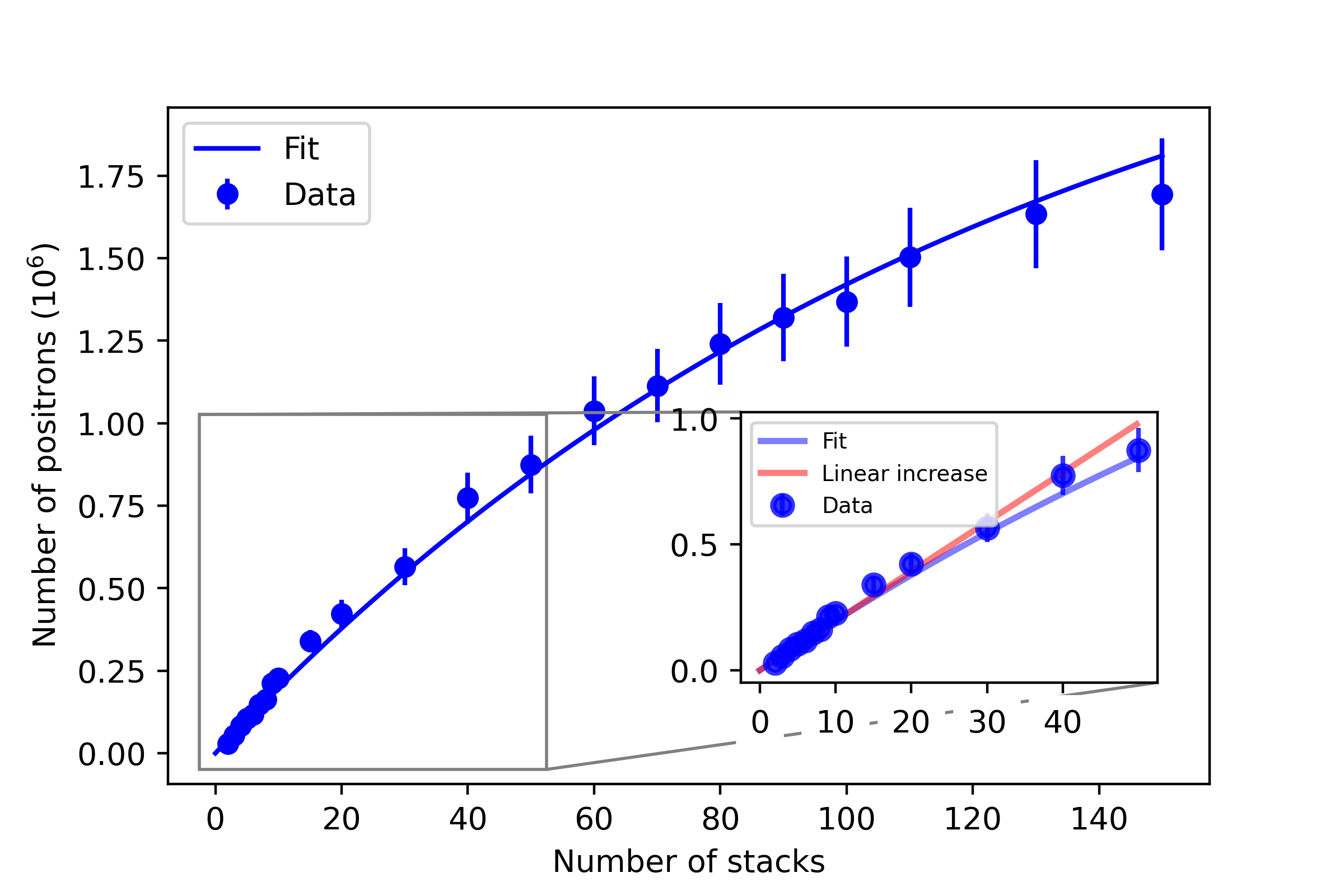}\caption{Number of positrons in the BGT (left) and accumulator (right) with the exponential rise to maximum fit (blue line). The insets show the deviation from a linear increase (red) in the number of positrons for low filling times/number of stacks .}
\label{fig:analysis}
\end{figure}

To calculate the lifetime of positrons in the BGT, it is filled for a variable time and the charge deposited on the MCP is measured with a charge amplifier. The data is fitted using an exponential rise to maximum, $\text{N(t)} = a\cdot\left[1-\text{exp}\left(-\text{t}/\tau\right)\right]$, where \textit{a} is the maximal number of positrons and $\tau$ the lifetime in the trap. The positrons have a lifetime of \unit[$(3.9 \pm 0.8)$]{s}.  Filling the BGT for \unit[1]{s} gives $(21300\pm2200)$ positrons, yielding a trapping efficiency of \unit[$(15\pm 4)$]{\%}. The left hand side plot in Fig.~\ref{fig:analysis} shows the measured number of positrons, the exponential rise to maximum fit, and the inset shows the deviation from the linear increase at low fill times.

The energy distribution of the positrons is measured by a retarding field analysis using the E1 electrode of the accumulator and looking at the annihilation signal with a plastic scintillator at the position indicated in Fig.~\ref{fig:experiment}. The resulting energy spread, i.e. the full width at half maximum of the energy distribution, is \unit[$(1.49\pm 0.01)$]{eV}, which is slightly smaller than the \unit[$(1.655\pm0.063)$]{eV} measured in Aarhus by \cite{Thomsen2010}. The size of the positron cloud is measured with a chevron stack MCP and the light produced in the attached phosphor screen is acquired with an Apogee ALTA-U77 CCD camera. The positrons are compressed with a quadrupolar rotating field with a frequency of \unit[6.5]{MHz} and an amplitude of \unit[0.5]{V}. The image is fitted by a 2-dimensional Gaussian yielding a radius of \unit[$(2.12\pm0.03)$]{mm} when the magnetic field ratios of the trap and the detector is taken into account. This result is about a factor of two larger then the results in \cite{Andersen2015}. The expansion rate of the positron cloud of \unit[$(2.24\pm0.08)$]{mm/s} is measured by stopping the RW drive and holding the positrons for a variable amount of time prior to extraction. 

\subsection{Accumulator}
The positron bunches are magnetically guided into the accumulator and are caught in a well formed between E1 and C9. Shortly after catching, they are moved to the downstream region and merged with positrons that were previously caught while applying a RW with frequency of \unit[100]{kHz} and an amplitude of \unit[10]{V}. The high pressure of \unit[$2.0\cdot10^{-7}$]{mbar} in the accumulator from the gases in the BGT limits the lifetime of the positrons to \unit[$(104\pm22)$]{s}. When the gas flow  into the trap is stopped by closing the upstream gate valve (GV1), then the positrons can be held longer: \unit[95]{\%} still remain after ten minutes. 

The MCP is mounted on a rotatable feedthrough, such that it can be turned to face either the BGT or the accumulator. For imaging or counting the charge deposited on the MCP, the positrons must be extracted upstream. This is a slightly different scheme than typically used for the transfer into the Cusp trap, which is performed by dumping from the downstream side of the accumulator. To achieve the same conditions for the positrons, they are moved to the same potential as if they were dumped towards the Cusp trap, but on the upstream side of the accumulator. The move has to be performed slowly to avoid positron losses or heating up the particles. By measuring the charge deposited on the MCP the transfer efficiency can be calculated. Due to the finite lifetime of positrons in the accumulator, the linear region of positron number vs. the stack number has been taken to evaluate the transfer efficiency (see inset right hand side of Fig.~\ref{fig:analysis}). From the slope in the linear region the number of positrons per stack is determined. This divided by the number of positrons delivered by the BGT in \unit[1]{s} yields the transfer efficiency. The resulting efficiency is  \unit[$92 ^{+8}_{-11}$]{\%}, which corresponds to $(19600 ^{+2700}_{-3100})$ positrons per transfer. The right hand side plot in Fig.~\ref{fig:analysis} shows the number of positrons as a function of number of stacks, the exponential rise to maximum fit, and the inset shows the deviation from a linear increase of the number of positrons for a low number of stacks.

Imaging of the extracted positrons yields a radius of \unit[$(1.44\pm0.05)$]{mm} to \unit[$(1.85\pm0.06)$]{mm} in the trap, depending on the number of stacks (\unit[5-160]{stacks}), and hence the number density of positrons. The expansion rate is measured using \unit[30]{stacks}, starting with a positron radius of \unit[$(1.63\pm0.04)$]{mm}, to be \unit[$(0.013\pm0.004)$]{mm/s} if GV1 is closed and \unit[$(0.045\pm0.003)$]{mm/s} if GV1 is open.

\section{Discussion}
\label{sec:discussion}
Table~\ref{tab:comparison} compares the efficiencies of the FPS system at CERN, the FPS system as described by \cite{Andersen2015}, the accumulator, and the system previously used by ASACUSA at CERN \citep{Murtagh2023}. The moderator efficiency of the previous system is \unit[$(0.25\pm0.1)$]{\%}, which is comparable with the new system. Andersen reported a moderator efficiency of \unit[0.5]{\%}, which would be a factor of two to three better. The reason for the smaller moderation efficiency achieved at CERN is under investigation. 

The trapping efficiency of the previous BGT is \unit[$(17.4\pm1.8)$]{\%}, which is very similar to the new system being \unit[$(15\pm4)$]{\%}. Andersen reported a trapping efficiency of \unit[35]{\%}. The smaller trapping efficiency at CERN may arise from a slight misalignment of the trap within the magnet, which was found by imaging the slow beam and comparing to images of the trapped positrons. It was also found that positrons with a larger radius were apertured only on one side, presumably by scraping the Gate electrode on their way to the detector. The misalignment may be addressed by using large correction coils around the trap magnet to produce a compensating transverse field.

The purpose of the accumulator is primarily to reduce the contamination of the vacuum of the Cusp trap during transfer. Prior to the transfer of the positrons, GV1 is closed, causing the pressure in the accumulator to fall from \unit[$2.0\cdot10^{-7}$]{mbar} to \unit[<$1\cdot10^{-8}$]{mbar} within three seconds. After that, the gate valves GV2 and GV3 between the positron system and the Cusp trap are opened. Previously, multiple bunches of positrons were accumulated in the Cusp trap, repeatedly opening GV2 and GV3 between the high pressure region from the positron trap (order of \unit[$10^{-6}$]{mbar}) and the UHV region of the Cusp trap ( \unit[<$10^{-12}$]{mbar}). Due to this repetitive transfer, and being unable to pump out the buffer gas in advance, the vacuum in the Cusp trap was contaminated, shortening the lifetime of the antiprotons and reducing the time available to produce antihydrogen. In the previous system the lifetime of the positrons in the BGT was \unit[40]{s}, which could significantly be increased to \unit[$(104\pm22)$]{s} (with cooling gas present) and \unit[>600]{s} (without cooling gas present) with the new accumulator. This increase in lifetime additionally means that a higher number of positrons can be sent into the Cusp trap per accumulation cycle.

\begin{table}
  \begin{center}
\def~{\hphantom{0}}
  \begin{tabular}{lcccc}
      Property  & FPS (CERN) & FPS (Aarhus) & Accumulator &  Old system \\
        Moderator efficiency (\%) & $0.2 \pm0.1$ & 0.5 & --   & $0.25 \pm 0.1$ \\  
        Trapping efficiency (\%) & $15\pm4$ & 35 & -- & $17.4\pm 1.8$ \\
        Transfer efficiency (\%) & -- & -- & $92^{+8}_{-11}$ & -- \\
        Lifetime  (s) & $3.9 \pm 0.8$ & 1.5 & $104\pm22$ & 40 \\

  \end{tabular}
  \caption{Comparison of the moderator efficiency, trapping efficiency and lifetime of the FPS system measured at CERN and at Aarhus University and the previous system. Additionally, the transfer efficiency and the lifetime of the accumulator are shown. }
  \label{tab:comparison}
  \end{center}
\end{table}

\section{Conclusions}
The RGM and the BGT have been installed in the ASACUSA experiment at CERN and recommissioned after the transport from Aarhus University. With a $89\,\mathrm{MBq}$ source, this system provides roughly $10^4$ positrons per second\textemdash comparable to the system it replaced. The positron bunches from the BGT are extracted once per second and accumulated in the newly developed and installed  accumulator. Up to \unit[40]{bunches} can be linearly stacked in the accumulator and held for more than ten minutes. The pressure in the accumulator decreases to <\unit[$10^{-8}$]{mbar} within three seconds after the flow is stopped by closing the upstream gate valve to the BGT. This low pressure prior to the transfer of positrons reduces the contamination of the vacuum in the Cusp trap, from which the ASACUSA experiment suffered previously.

With the current system it is possible to accumulate $1.5\cdot 10^6$ positrons in \unit[110]{s}. To increase the number of positrons from the RGM a new $^{22}$Na source with an activity of \unit[1.89]{GBq} was ordered. With the presented moderator, trapping and transfer efficiencies, the number of positrons per second is expected to increase by a factor of 20. 

The increased lifetime of positrons in the accumulator allows for a longer accumulation time. This increases the number of positrons which can be transferred into the Cusp trap in a single positron transfer cycle. The ASACUSA-Cusp collaboration optimised the antiproton preparation cycle last year, such that each antiproton extraction from CERN every \unit[108]{s} can be used. The lifetime of \unit[($104\pm22$)]{s} in the accumulator means that (with the new source) only one transfer cycle should be necessary for antihydrogen production. 

To further increase the lifetime in the accumulator during filling, the installation of a pumping restriction downstream of the BGT is under consideration. Future work will also focus on the moderator efficiency, which is currently a factor of two to three lower than previously achieved at Aarhus University. The reason for this reduced efficiency is not yet understood, however investigations are ongoing. When the cause for the reduction in trapping efficiency and moderator efficiency are found, the upgraded system would not only have been successful in minimising the contamination of the Cusp trap, but also increased the efficiency of positron preparation by a factor of four compared to the previous system.

\section*{Acknowledgments}
This work was supported by the Austrian Science Fund (FWF) Grant Nos. P 32468, W1252-N27, and P 34438; the JSPS KAKENHI Fostering Joint International Research Grant No. B 19KK0075; the Grant-in-Aid for Scientific Research Grant No. B 20H01930; Special Research Projects for Basic Science of RIKEN; Università di Brescia and Istituto Nazionale di Fisica Nucleare; and the European Union's Horizon 2020 research and innovation program under the Marie Sklodowska-Curie Grant Agreement No. 721559.

\section*{Declaration of interests}
The authors report no conflict of interest.

\bibliographystyle{jpp}
\bibliography{Bibliography}

\end{document}